\documentclass[twocolumn,english,preprintnumbers,amsmath,amssymb]{revtex4}
\usepackage[T1]{fontenc}
\usepackage[latin1]{inputenc}
\usepackage{color}
\usepackage{graphicx}

\makeatletter

\newcommand{\noun}[1]{\textsc{#1}}
\usepackage{graphicx}

\usepackage{graphicx}

\usepackage{graphicx}% Include figure files
\usepackage{dcolumn}% Align table columns on decimal point
\usepackage{bm}% bold math

\usepackage{babel}

\usepackage{babel}

\usepackage{babel}
\makeatother
\begin{document}

\title{\noun{Physics Models of Earthquake}}

\author{Srutarshi Pradhan}

\email{pradhan.srutarshi@ntnu.no}

\address{Department of Physics, Norwegian University of Science and Technology,
N--7491 Trondheim, Norway}

\begin{abstract}
Since long back, scientists have been putting enormous effort to understand
earthquake dynamics -the goal is to develop a successful prediction
scheme which can provide reliable alarm that an earthquake is imminent.
Model studies sometimes help to understand in some extend the basic
dynamics of the real systems and therefore is an important part of
earthquake research. In this report, we review several physics models
which capture some essential features of earthquake phenomenon and
also suggest methods to predict catastrophic events being within the
range of model parameters. 
\end{abstract}
\maketitle
\begin{flushleft}\textbf{\emph{Introduction}}\par\end{flushleft}

\vskip.1in

\noindent `Earthquake' is a long standing problem to the scientist
community as it is a subject of many unknowns. Strong earthquakes
often causes huge devastation in terms of lives and properties and
therefore this subject always demands high priority. But so far, little
advancement has been made to understand the entire earthquake-dynamics
and the challenge still remains -to develop a successful prediction
scheme which can provide exact space-time information of future earthquakes
and their expected magnitudes. However, several models and hypothesis
\cite{book-1,book-2,book-3,book-4,GR54,PB06} have been proposed and
some recent studies suggest potential methods to predict catastrophic
events in some earthquake models. In this short report we discuss
such physics models of earthquake, giving importance to their capability
of prediction.

\vskip.1in

\begin{flushleft}\textbf{\emph{Geological facts}}\par\end{flushleft}

\vskip.1in

Plate-tectonic theory explains the origin of earthquake through a
stick-slip dynamics: Earth's solid outer crust (about 20 km thick)
rests on a tectonic shell which is divided into numbers (about 12)
of mobile plates, having relative velocities of the order of few centimeters
per year. This motion of the plates arises due to the powerful convective
flow of the earth's mantle at the inner core of earth. On the other
hand solid-solid frictional force arises at the crust-plate boundary
and it sticks them together. This kind of sticking develops elastic
strains and the strain energy gradually increases because of the uniform
motion of the tectonic plates. Therefore a competition comes to play
between the sticking frictional force and the restoring elastic force
(stress). When the accumulated stress exceeds the frictional force,
a slip (earthquake) occurs and it releases the stored elastic energy
in the form of sound, heat and mechanical vibrations. It has been
observed that generally a series of small earthquakes appear before
(foreshocks) and after (aftershocks) a big quake (main shock). 

\begin{center}\includegraphics[width=3in,height=2in]{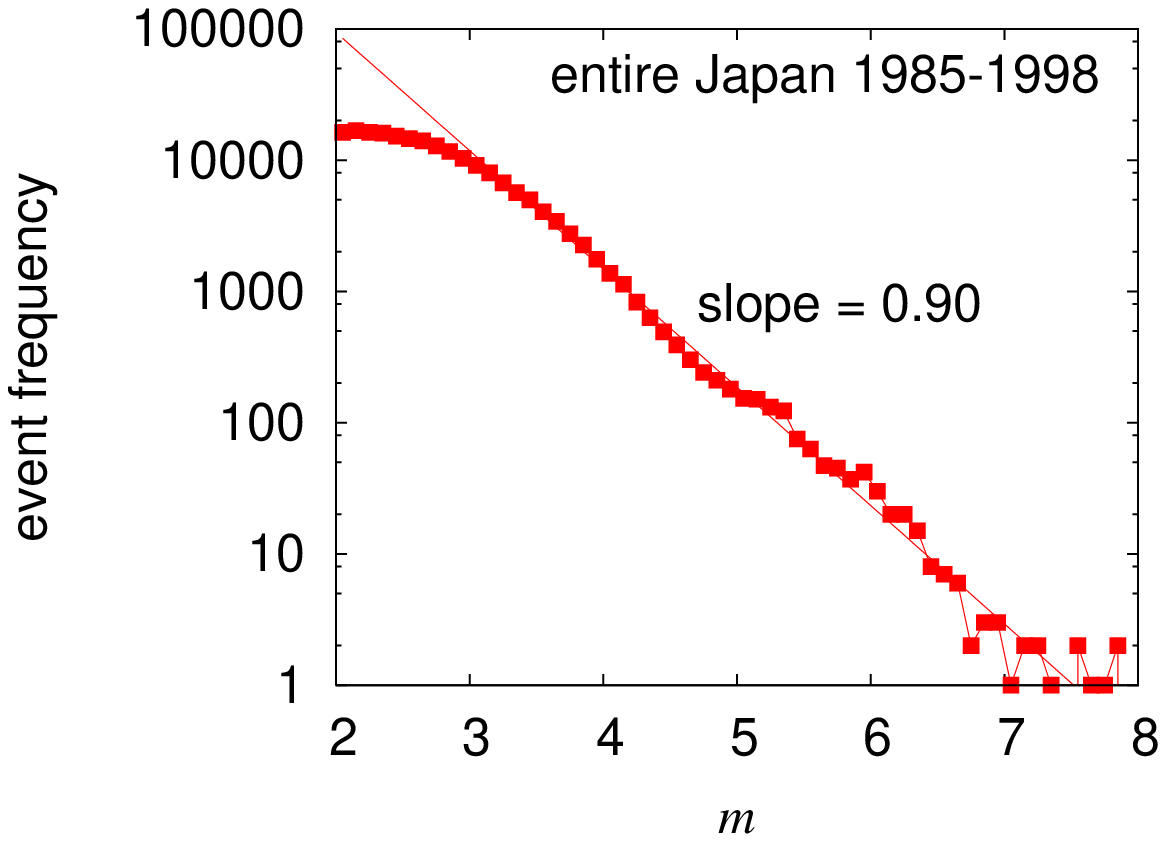}\par\end{center}

{\footnotesize Figure.} \textbf{\footnotesize 1}{\footnotesize :}
\textbf{\footnotesize }{\footnotesize The magnitude distribution of
earthquakes in Japan (from JUNEC catalog) \cite{Kawamura-06}. The
straight line with slope $-0.9$ is the best fit to the data points
for $m>3$ and supports Gutenberg-Richter law. }{\footnotesize \par}

\vskip.2in

\begin{flushleft}\textbf{\emph{Physics models}}\par\end{flushleft}

\vskip.1in

The overall frequency distribution of earthquakes including foreshocks,
mainshocks and aftershocks, seems to follow the empirical Gutenberg-Richter
law \cite{GR54}: \[
\ln N(m)=constant-bm,\]
where $N(m)$ is the number of earthquakes having magnitude (in Richter
scale) greater than or equal to $m$, $b$ is the power exponent.
The observed value of $b$ ranges between $0.7$ and $1.0$ (Fig.1).
The amount of energy $\epsilon$ released in an earthquake is related
to the magnitude as \[
\ln\epsilon=constant+am.\]
Therefore Gutenberg-Richter law can be expressed as an alternative
form: \[
N(\epsilon)=\epsilon^{-\alpha},\]
where $N(\epsilon)$ is the number of earthquakes releasing energy
greater than or equal to $\epsilon$ and $\alpha=b/a$. The Gutenberg-Richter
law is being considered as one of the most fundamental observations
by the physicists.

Several models have been proposed to study the nature of the earthquake
phenomenon. The main intension is to capture the Gutenberg-Richter
type power law for the frequency distribution of failures (quakes)
by modelling different aspects of faults --structure, material properties,
geometry etc. These models can be classified into four groups according
to their basic assumptions: (A) Friction models, incorporate the stick-slip
dynamics through the collective motion of an assembly of locally connected
elements subject to slow driving force (B) Fracture models, look at
earthquake phenomena as a fracture-failure process of deformable materials
that break under external loading through slow build-up of stress
(C) Self-organised critical (SOC) models, consider earthquake as a
self-driven slow process and (D) Fractal models, give importance to
fractal nature of the crust-plate interfaces and address the phenomenon
as a two-fractal overlap problem. 

\vskip.1in

\begin{flushleft}\textbf{(A)} \textbf{\emph{Friction models}}\par\end{flushleft}

\vskip.1in

In 1967 Burridge and Knopoff \cite{BK67} introduced model studies
in earthquake research. They proposed a spring-block model to mimic
the typical stick-slip dynamics of earthquake phenomena, which has
been extended later by Carlson and Langer \cite{CL89}. The Burridge-Knopoff
type model contains a linear array of blocks of mass $m$ coupled
to each other by identical harmonic springs of strength $k_{c}$ and
also attached to a fixed surface at the top by a different set of
identical springs having strength $k_{p}$. The blocks are kept on
a horizontal platform (rough surface) which moves with a uniform velocity
$V$ (Fig.2). Here qualitatively the blocks can be thought of as the
points of contact between two plates moving at a relative speed $V$,
where the spring constants $k_{c}$ and $k_{p}$ represents the linear
elastic response of the contact region to compression and shear respectively. 

\begin{center}\includegraphics[%
  width=3in,
  height=1.5in]{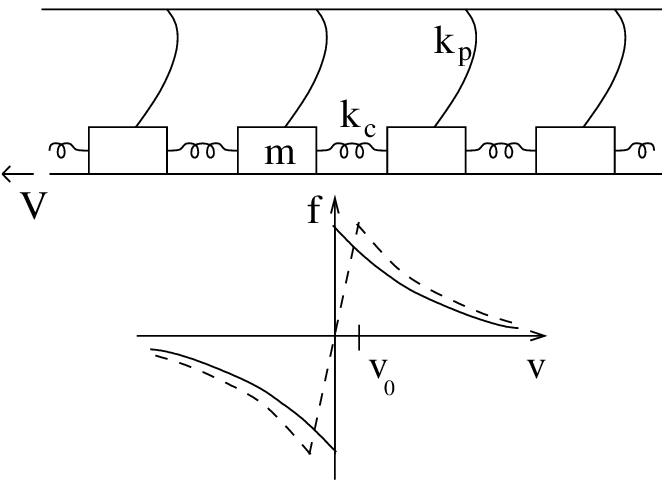}\par\end{center}

{\footnotesize Figure} \textbf{\footnotesize 2}{\footnotesize :} \textbf{\footnotesize }{\footnotesize The
Burridge-Knopoff model (above) and two different forms of velocity
weakening friction force (below) \cite{rumi}.}{\footnotesize \par}

\vskip.1in

\noindent Starting from unstrained condition, when the system is pulled
slowly (at constant rate), the blocks initially remain stuck to the
surface due to friction between surfaces. A slip of a block occurs
when the corresponding spring force overcomes the threshold value
(maximum static frictional force between that block and the rough
surface). The dynamics of each block is basically the resultant of
two phases: a static phase and a dynamic phase. During static phase
the block remains stuck on the rough surface and the elastic strain
continuously grows up. The static phase comes to a sudden end when
driving force on that block attains the threshold value and dynamic
phase begins. Due to the presence of frictional force this dynamic
phase is definitely dissipative in nature. This dissipation reduces
the relative velocity of the block and the system again goes back
to the static phase. The dynamic friction is assumed as a velocity
weakening function $f$ (shown in the Fig.2 ). Presence of spring
force and velocity weakening friction force leads the system to a
complex dynamical state. At the initial stage, individual small slip
occurs. But due to constant pulling, the springs are gradually stretched
and attain the limit of total frictional stability of the blocks where
the collective slips of almost all the blocks occur. Clearly this
is a big event and a large amount of elastic energy is released here.
The equation of motion of the $j^{th}$ block of the system is \[
m\ddot{x}_{j}=k_{c}(x_{j+1}-2x_{j}+x_{j-1})-k_{p}x_{j}-f(\dot{x}_{j}-V)\]
 where dots denote differentiation with respect to $t$, $m$ is the
mass of a block, $k_{p}$ and $k_{c}$ are spring constants of the
connecting springs and $f$ represents the nonlinear velocity weakening
friction force. The position coordinate $x$ is measured along the
chain length and the velocity $V$ of the platform is in the increasing
$x$ direction. The total energy of the system at time $t$ can be
calculated by solving the above equation. A sudden drops in the energy
of the system is identified as the released energy during a slip event
and the distribution of such energies follows power-law: \[
N(\epsilon)\sim\epsilon^{-c},c\sim1,\]
 where $N(\epsilon)$ is the number of slips releasing energy greater
than or equal to $\epsilon$. 

\vskip.1in

\begin{flushleft}\emph{Prediction possibility of major events (slips) }\par\end{flushleft}

Recently Dey et al. \cite{rumi} developed a method to predict the
major slip event in Burridge-Knopoff type spring-block model. Introducing
an additional dissipative force in the spring-block arrangement they
identified the dissipative functional $R(t)$ as energy bursts similar
to the acoustic emission signals \cite{AE-1,AE-2} observed in experiments.
The distribution of $R(t)$ shows power laws if one records all slip
events including the major slips.

\begin{center}\includegraphics[width=2.5in,height=1.5in]{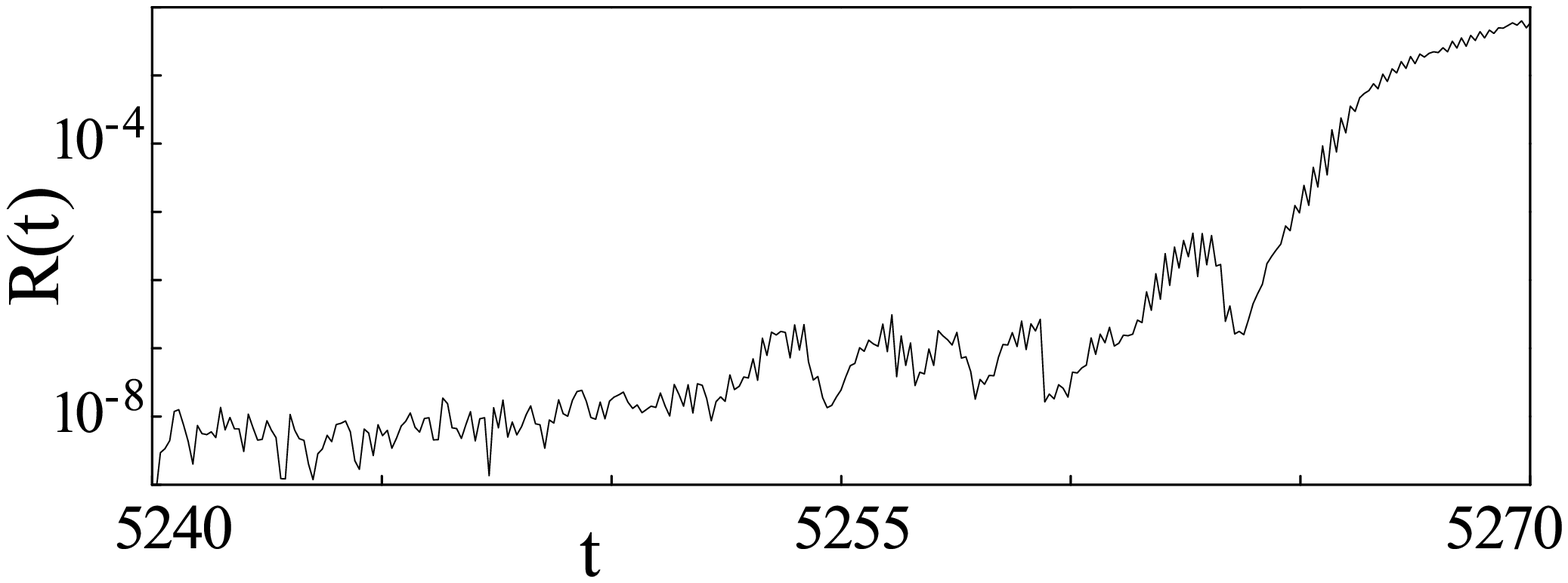}\par\end{center}

{\footnotesize Figure} \textbf{\footnotesize 3}{\footnotesize :} \textbf{\footnotesize }{\footnotesize The
dissipative functional $R(t)$ vs. time ($t$) in Burridge-Knopoff
model \cite{rumi}.}{\footnotesize \par}

\vskip.1in

The plot of $R(t)$ vs. $t$ shows a gradual increase in activity
(Fig.3) prior to the occurrence of a major slip. But as $R(t)$ is
noisy it can not help much to predict the major slip event, rather
the cumulative energy dissipated $E_{ae}(t)\sim\int_{0}^{t}R(t^{'})dt^{'}$
grows in steps and it seems to diverge as a major slip event is approached
(Fig.4). From such divergence one can predict the occurrence time
($t_{c}$) of a major slip through proper extrapolation. 

\begin{center}\includegraphics[width=2.5in,height=1.8in]{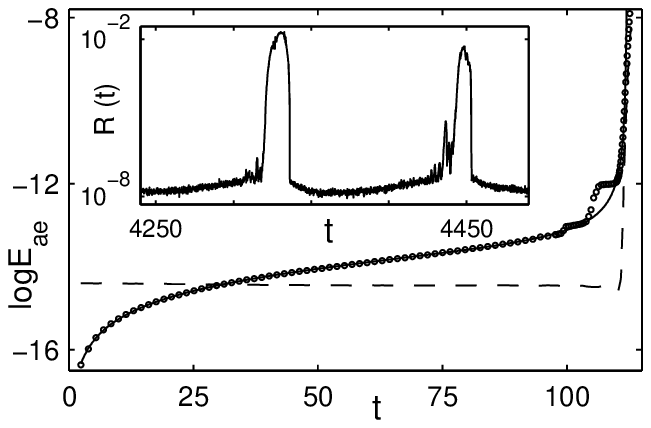}\par\end{center}

{\footnotesize Figure} \textbf{\footnotesize 4}{\footnotesize :} \textbf{\footnotesize }{\footnotesize Cumulative
energy $E_{ae}$ versus time ($t$) plot. Inset shows the time series
of $R(t)$ including two major slip events \cite{rumi}. }{\footnotesize \par}

\vskip.1in

\begin{flushleft}\textbf{(B)} \textbf{\emph{Fracture models}}\par\end{flushleft}

\vskip.1in

Earthquake can be considered as a fracture-failure phenomenon through
slow build up of stress at the plate-crust boundary with the movement
of the plate as external driving force. The deformation properties
of the materials sitting at the boundary play crucial role on spatial
redistribution of the stress around a broken region and this actually
decides in which direction the crack front should propagate. Fiber
bundle model and random fuse model address such scenario in material
breakdown and sometimes they are treated as models of earthquake since
they produce time series of avalanches which follow power law distribution
similar to Gutenberg-Richter law.

\subsection*{\textmd{\emph{Fiber bundle model}}}

\noindent Fiber bundle (RFB) model consists of many ($N_{0}$) fibers
connected in parallel to each other and clamped at their two ends
and having randomly distributed strengths. The model exhibits a typical
relaxational dynamics when external load is applied uniformly at the
bottom end (Fig.5). In the global load-sharing approximation \cite{FT,Dan,HH-92},
surviving fibers share equally the external load. Initially, after
the load $F$ is applied on the bundle, fibers having strength less
than the applied stress $\sigma=F/N_{0}$ fail immediately. After
this, the total load on the bundle redistributes globally as the stress
is transferred from broken fibers to the remaining unbroken ones.
This redistribution causes secondary failures which in general causes
further failures and produces an {}``avalanche'' which denotes simultaneous
failure of several elements. With steady increase of external load,
avalanches of different size appear before the global breakdown where
the bundle collapses. The scaling properties of such mean-field dynamics
and the avalanche statistics are expected to be extremely useful in
analysing fracture and breakdown in real materials, including earthquakes
\cite{D92,More94,KHH-97,JV-97}.

\begin{center}\includegraphics[%
  width=1.5in,
  height=1.3in]{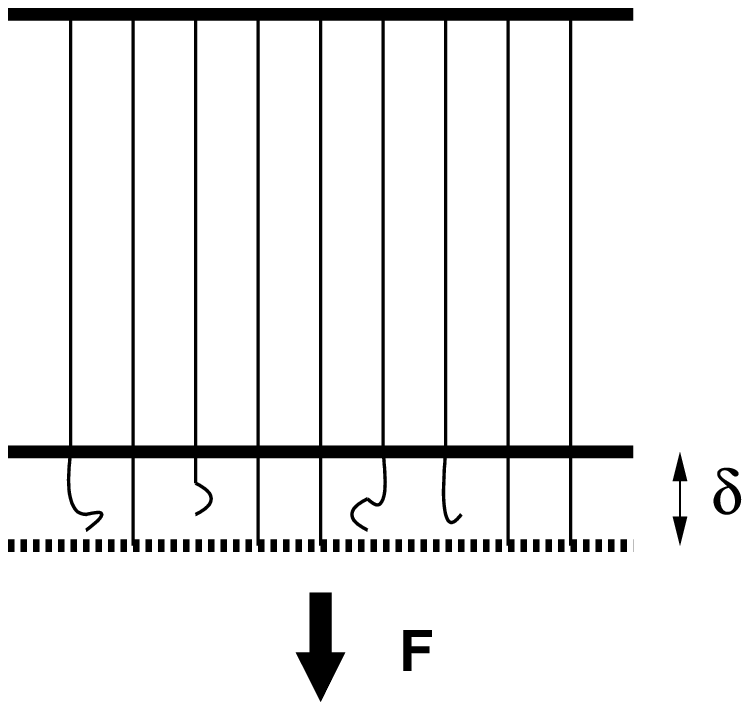}\par\end{center}

{\footnotesize Figure} \textbf{\footnotesize 5}{\footnotesize :} \textbf{\footnotesize }{\footnotesize The
fiber bundle model . }{\footnotesize \par}

\vskip.1in

When external load is applied, the surviving fraction of total fibers
follows a simple recursion relation \[
U_{t+1}(\sigma)=1-P(\sigma/U_{t}),\]

\noindent where $U_{t}=N_{t}/N_{0}$, $\sigma$ is the external stress
and $P$ is cumulative probability function. The recursion relation
has the form of an iterative map $U_{t+1}=Y(U_{t})$ and finally the
dynamics stops at a fixed point where $U_{t+1}=U_{t}$.

If external load is increased in steps by equal amount $\Delta F$,
then the entire failure process can be formulated through the recursive
dynamics \cite{SPB-02} mentioned above and the fixed point solution
gives the value of critical stress $\sigma_{c}$ above which the bundle
collapses. It can be shown that the bundle undergoes a phase transition
from partially broken state to completely broken state. The order
parameter ($O$), susceptibility ($\chi$) and relaxation time ($\tau$)
follow robust power laws with universal exponent values \cite{PSB-03}. 

In case of quasi-static load increment, only the weakest fiber (among
the intact fibers) fails after loading and then the bundle undergoes
load redistribution till a fixed point is reached. The fluctuations
in strength distributions produces different size of avalanches during
the entire failure process and the avalanche distribution follows
power law (Fig.6) with exponent $-5/2$. Hemmer and Hansen \cite{HH-92}
has analytical proved that this exponent is universal under mild restriction
on strength distributions. 

\begin{center}\includegraphics[%
  width=1.5in,
  height=1.5in]{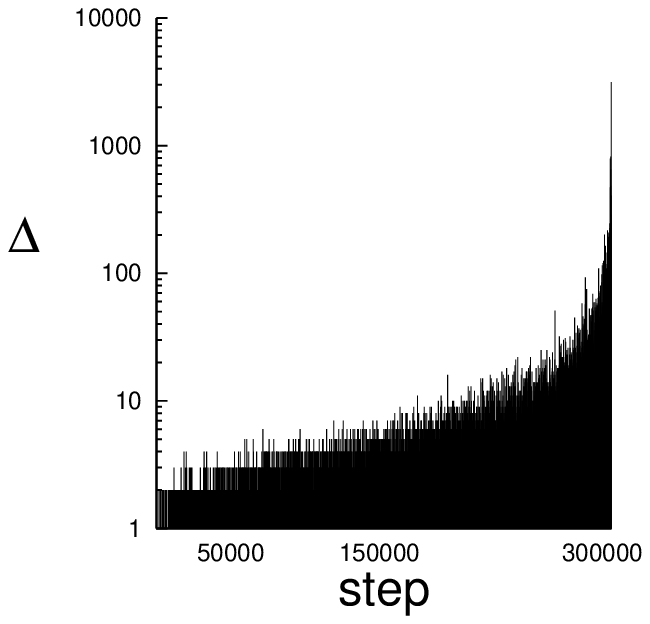}\includegraphics[%
  width=1.5in,
  height=1.5in]{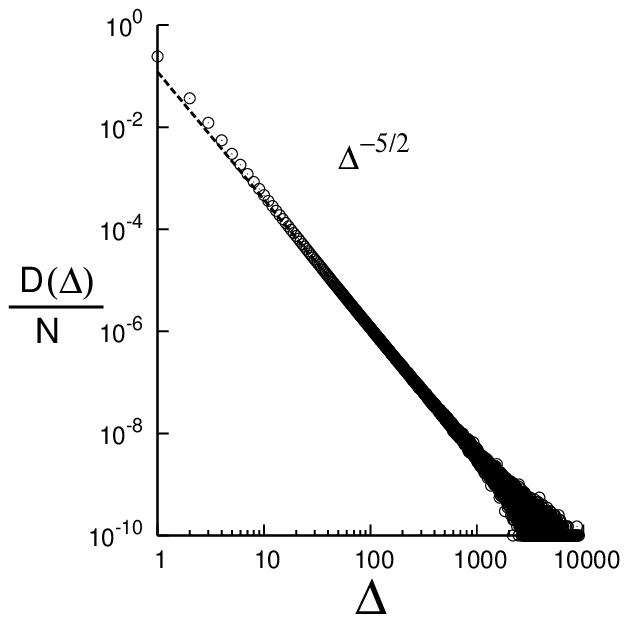}\par\end{center}

{\footnotesize Figure} \textbf{\footnotesize 6}{\footnotesize :} \textbf{\footnotesize }{\footnotesize Avalanche
time series in a fiber bundle model (left) and the corresponding avalanche
distribution (right). }{\footnotesize \par}

\vskip.2in

\subsection*{\textmd{\emph{Random fuse model}}}

\begin{center}\includegraphics[%
  width=1.5in,
  height=1.2in]{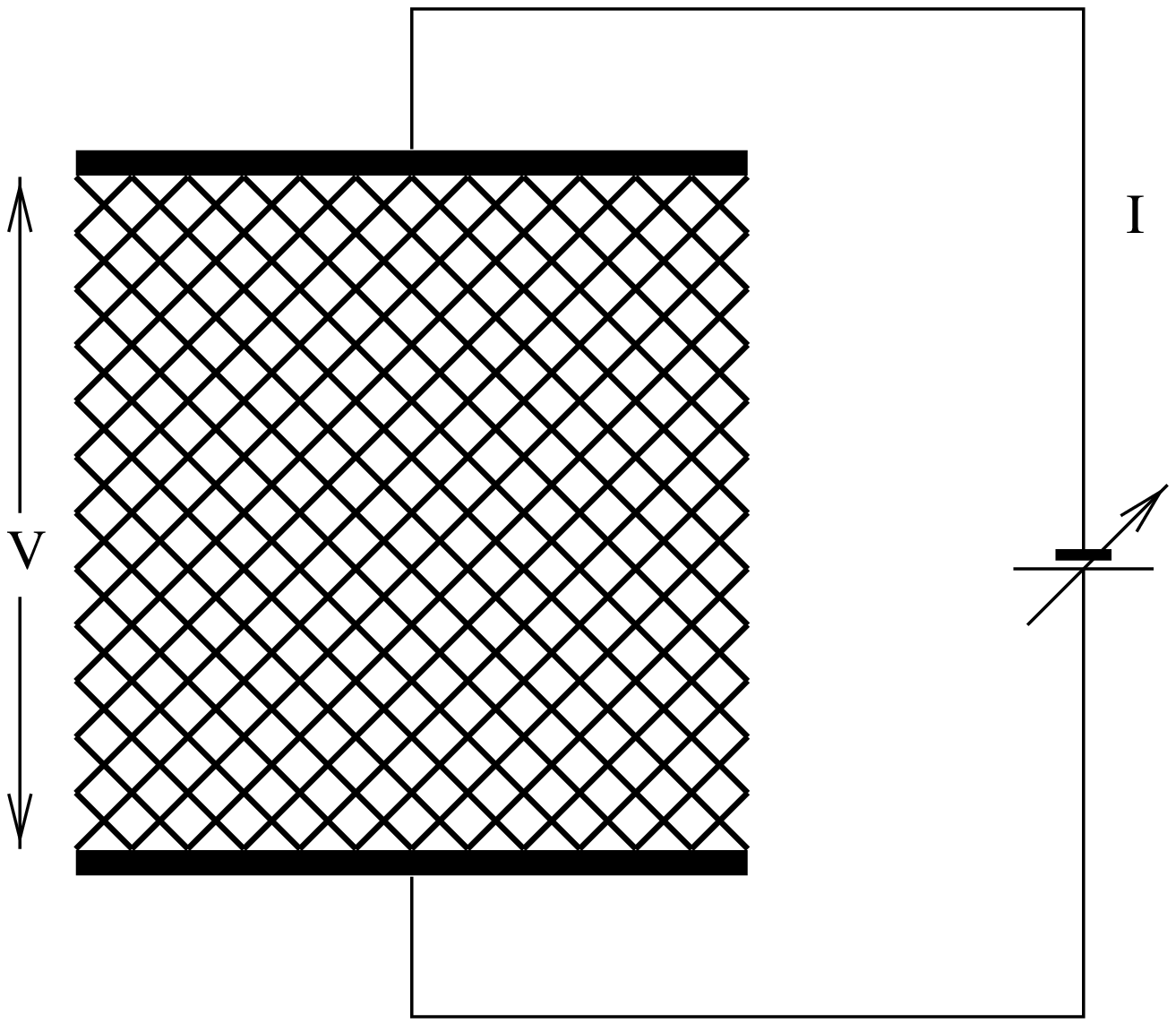}\par\end{center}

{\footnotesize Figure} \textbf{\footnotesize 7}{\footnotesize :} \textbf{\footnotesize }{\footnotesize The
random fuse model. }{\footnotesize \par}

\vskip.1in

The fuse model \emph{}\cite{fuse-94,PS97,fuse-05} consists of a lattice
in which each bond is a fuse, i.e., an ohmic resistor as long as the
electric current it carries is below a threshold value. If the threshold
is passed, the fuse burns out irreversibly. The threshold $t$ of
each bond is drawn from an uncorrelated distribution $p(t)$. The
lattice is placed at $45{}^{\circ}$ with regards to the electrical
bus bars (Fig.7) and an increasing current is passed through it. Numerically,
the Kirchhoff equations are solved at each node. 

When a bond breaks, current value on the neighboring bonds increases
and sometimes it triggers secondary failures. Finally the system comes
to a state where no current passes through the lattice; that means
there is a crack which separates the lattice in two pieces. With gradual
increase of current/voltage a series of intermediate avalanches appear
before the final breakdown. The distribution of such avalanches follows
power law with exponent close to $3$ (Fig.8).

\begin{center}\includegraphics[%
  width=1.5in,
  height=1.5in]{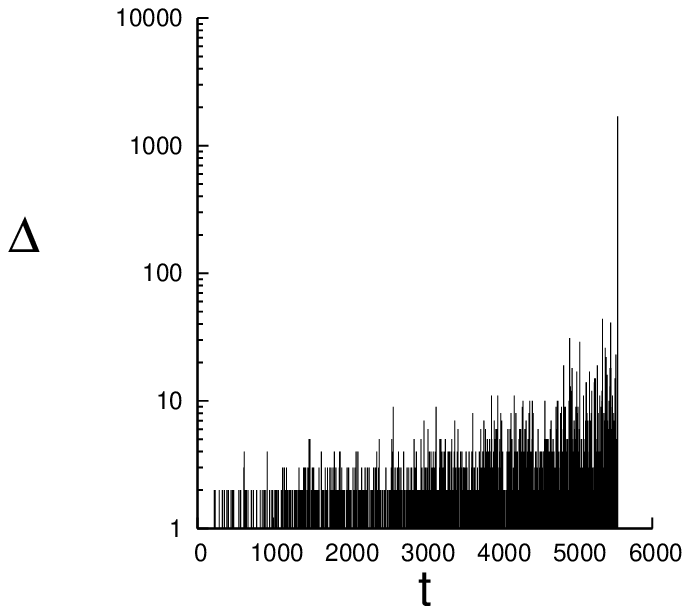}\includegraphics[%
  width=1.5in,
  height=1.5in]{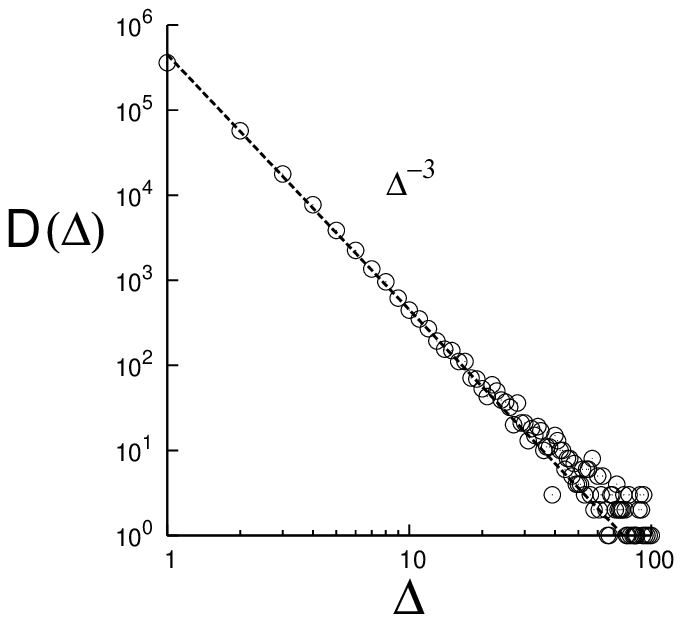}\par\end{center}

{\footnotesize Figure} \textbf{\footnotesize 8}{\footnotesize :} \textbf{\footnotesize }{\footnotesize Avalanche
time series in a fuse model (left) and the corresponding avalanche
distribution (right). }{\footnotesize \par}

\vskip.1in

\subsubsection*{\textmd{Crossover behavior: signature of imminent breakdown}}

A robust crossover behavior has been observed \cite{PH-04,PHH-05,PHH-06}
recently in the two very different models described above where the
system gradually approaches the global failure through several intermediate
failure events. If intermediate avalanches are recorded, avalanche
distribution follows a power law with an exponent that crosses over
from one value to a very different value when the system is close
to the global failure or breakdown point (Fig.9). Therefore, this
crossover is a signature of imminent breakdown. 

\begin{center}\includegraphics[%
  width=1.5in,
  height=1.5in]{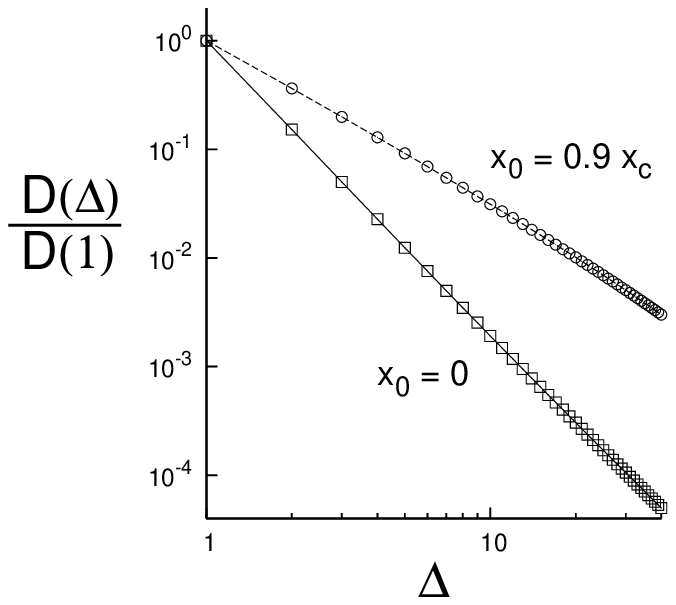}\includegraphics[%
  width=1.5in,
  height=1.5in]{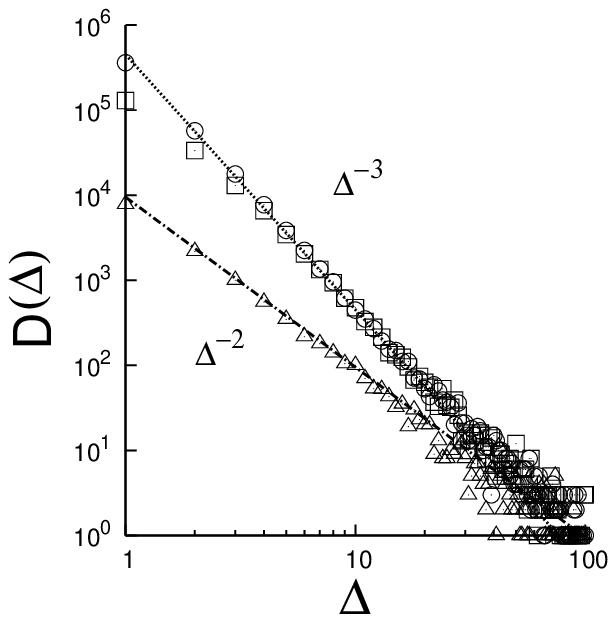}\par\end{center}

{\footnotesize Figure} \textbf{\footnotesize 9}{\footnotesize :} \textbf{\footnotesize }{\footnotesize Crossover
signature in avalanche power law. In fiber bundle model: $x_{0}$
is the starting position of recording avalanches and $x_{c}$ is the
global failure point; exponent value changes from $5/2$ to $3/2$
(left) and in random fuse model: exponent value changes from $3$
to $2$ when the system comes closer to breakdown point (right). }{\footnotesize \par}

\vskip.1in

Recently Kawamura \cite{Kawamura-06} observed similar crossover behavior
for the local magnitude distribution of earthquakes in Japan (Fig.10).
This observation has strengthened the possibility of using crossover
signal as a tool of predicting catastrophic events. 

\begin{center}\includegraphics[width=2in,height=1.8in]{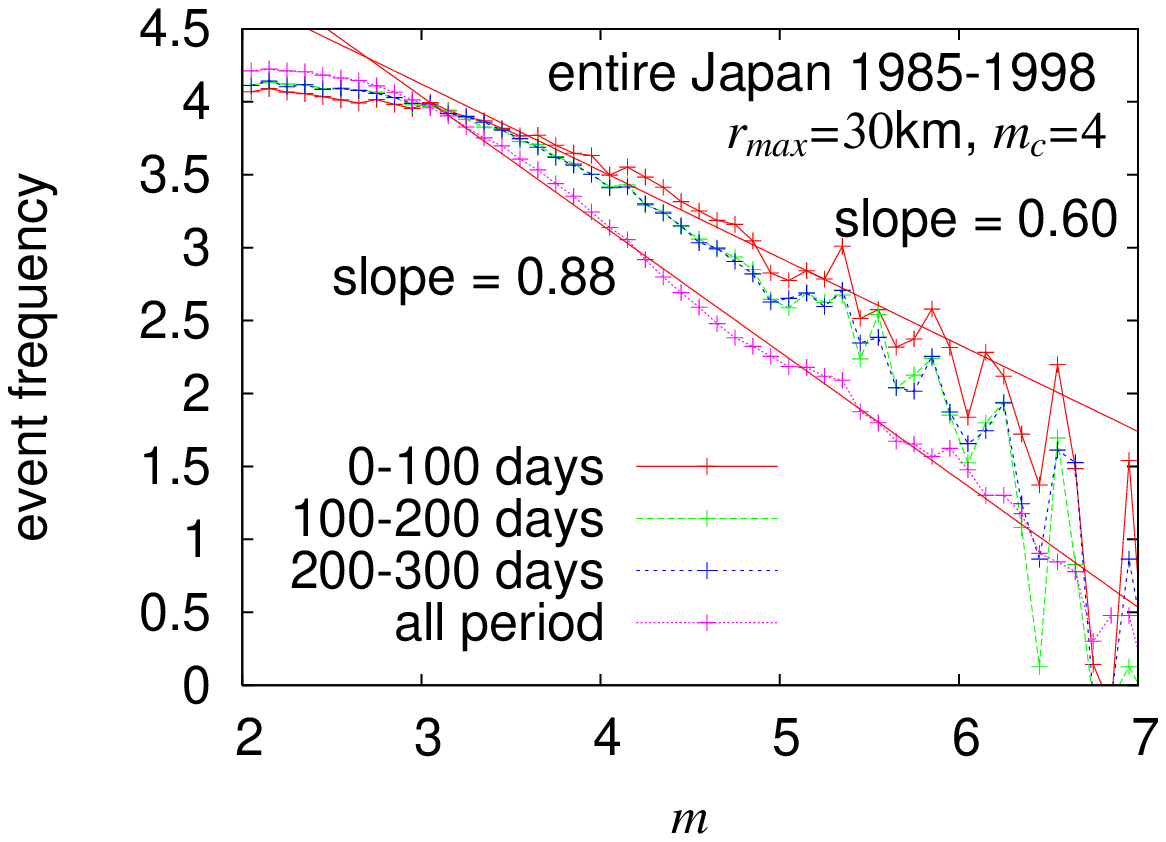}\par\end{center}

{\footnotesize Figure} \textbf{\footnotesize 10}{\footnotesize :}
\textbf{\footnotesize }{\footnotesize Crossover signature in the magnitude
distribution of earthquakes within Japan \cite{Kawamura-06}.}{\footnotesize \par}

\vskip.2in

\begin{flushleft}\textbf{(C)} \textbf{\emph{Self-organised critical
models}}\par\end{flushleft}

\vskip.1in

In various thermodynamics systems, there is a {}``critical'' point
where the systems become totally correlated and show scale free (power
law) behaviour. Apart from the critical point the average microscopic
quantities of the systems follow scaling behaviour. Generally such
critical states are achieved through the fine-tuning of physical parameters,
such as temperature, pressure etc. and the power law behavior is considered
to be a signature of the {}``critical'' state of the system. However,
it has been observed that some complex systems evolve collectively
to such critical state only through mutual interactions and show power
law behavior there. These systems do not need any fine tuning of physical
parameter and therefore are considered as {}``self-organised critical''
(SOC) systems. The term SOC was first introduced by Bak, Tang and
Wiesenfeld in 1987.

The magnitude distribution of earthquake shows power-law (Gutenberg-Richter
law), therefore it is tempting to assume that earthquake happens through
a self-organised dynamics: the build up of stress due to tectonic
motion is a self organised slow process; gradually the critical state
is achieved where the stress releases in bursts of various sizes.
Extensive research have been going on to establish relation between
earthquake and SOC systems, for which several models have been proposed.
So far, sandpile models are the best example of SOC system.

\subsection*{\textmd{\emph{BTW sandpile model}}}

The first attempt to study SOC through model systems was made by Bak,
Tang and Wiesenfeld \cite{BTW}. This is a model of sandpile whose
natural dynamics drives it towards the critical state. The model can
be described on a two dimensional square lattice. At each lattice
site $(i,j)$, there is an integer variable $h_{i,j}$ which represents
the height of the sand column at that site. A unit of height (one
sand grain) is added at a randomly chosen site at each time step and
the system evolves in discrete time. The dynamics starts as soon as
any site $(i,j)$ has got a height equal to the threshold value ($h_{th}$=
$4$): that site topples, i.e., $h_{i,j}$ becomes zero there, and
the heights of the four neighbouring sites increase by one unit \[
h_{i,j}\rightarrow h_{i,j}-4,h_{i\pm1,j}\rightarrow h_{i\pm1,j}+1,h_{i,j\pm1}\rightarrow h_{i,j\pm1}+1.\]

\begin{center}\includegraphics[%
  width=3in,
  height=1in]{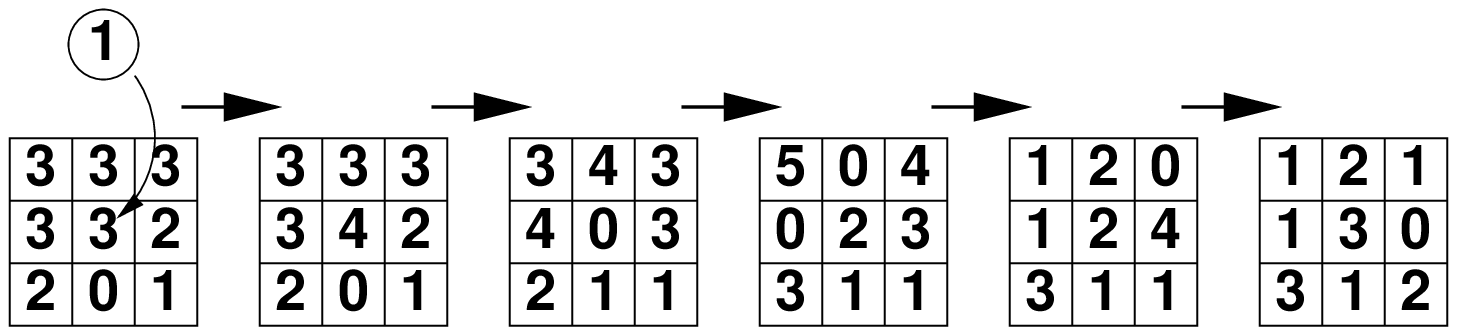}\par\end{center}

{\footnotesize Figure} \textbf{\footnotesize 11}{\footnotesize :}
\textbf{\footnotesize }{\footnotesize BTW model on square lattice
\cite{manna}.}{\footnotesize \par}

\vskip.1in

The process continues till all sites become stable (Fig.11). In case
of toppling at the boundary of the lattice ($4$ nearest neighbours
are not available), grains falling outside the lattice are removed
and considered to be absorbed/collected at the boundary. Gradually
the average height $h_{av}$ attains a critical value $h_{c}$, beyond
which it does not increase at all - on an average the additional sand
grains are flown away from the lattice . Total number of toppling
between two successive stable states, determines the size of an avalanche
and at the critical state avalanche size distribution follows power
laws: $n_{s}\sim s^{-\Gamma}$, where $n_{s}$ denotes the density
of $s$ size avalanches. The exponent $\Gamma$ has the value $\Gamma\simeq1.15\pm0.10$
in 2D \cite{manna}.

\subsection*{\textmd{\emph{Manna model}}}

Manna proposed the stochastic sand-pile model \cite{manna} by introducing
randomness in the dynamics of sand-pile growth. Here, the critical
height is $2$. Therefore at each toppling, the two rejected grains
choose their host among the four available neighbours randomly with
equal probability. After constant adding of sand grains, the system
ultimately settles at a critical state having height $h_{c}$ and
exhibits scale free behavior in terms of avalanche and life time distributions.
But the power law exponents are different compared to those in BTW
model and therefore Manna model belongs to a different universality
class \cite{dhar}. 

\begin{center}\includegraphics[width=1.7in,height=1.8in]{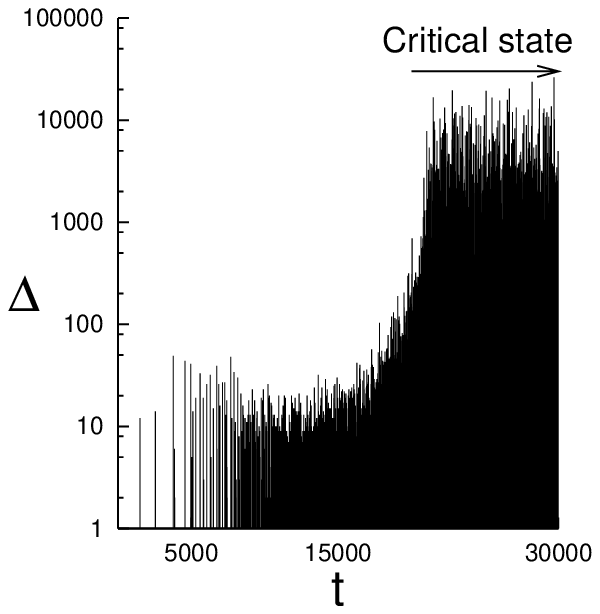}\includegraphics[width=1.7in,height=1.8in]{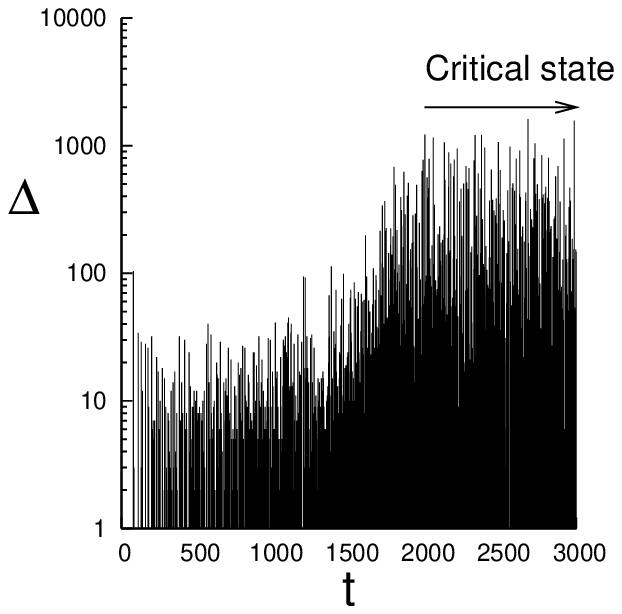}\par\end{center}

{\footnotesize Figure} \textbf{\footnotesize 12}{\footnotesize :}
\textbf{\footnotesize }{\footnotesize Avalanche time series in a BTW
model (left) and in a Manna model (right). }{\footnotesize \par}

\subsection*{\textmd{\emph{Precursory activities }}}

On constant adding of grains, sandpile models gradually attains critical
state from sub-critical states. Now the question is: can one predict
the critical state from the sub-critical response of the pile? A pulse
perturbation method \cite{Acharyya,SB01} gives the answer:

At an average height $h_{av}$, a fixed number of height units $h_{p}$
(pulse of sand grains) is added at any central point of the system.
Just after this addition, the local dynamics starts and it takes a
finite time or iterations to return back to the stable state after
several toppling events. \textcolor{black}{One can measure the response
parameters: $\Delta$ $\rightarrow$ number of toppling $\tau$ $\rightarrow$
number of iteration and $\xi$ $\rightarrow$ correlation length}
which is the distance of the furthest toppled site from the site where
$h_{p}$ has been dropped.

\textcolor{black}{To ensure toppling at the target site, the pulse
hight has been chosen as $h_{p}=4$ for BTW model and $h_{p}=2$ for
Manna model. Simulation studies conclude \cite{SB01} that all the
response parameter follow power law as $h_{c}$ is approached: $\Delta\propto(h_{c}-h_{av})$$^{-\lambda}$,
$\tau\propto(h_{c}-h_{av})^{-\mu}$, $\xi\propto(h_{c}-h_{av})^{-\nu}$;
$\lambda\cong2.0$, $\mu\cong1.2$ and $\nu\cong1.0$. Now if $\Delta^{-1/\lambda}$,
$\tau^{-1/\mu}$ and $\xi^{-1/\nu}$ are plotted against $h_{av}$,
all the curve follow straight line and they should touch the x axis
at $h_{av}=h_{c}$. A proper extrapolation estimates the critical
hight $h_{c}$ $=2.13\pm.01$ for BTW model and $h_{c}$ $=0.72\pm.01$
for Manna model, which agree well with direct estimates \cite{manna}.}

\begin{flushleft}\includegraphics[width=2.2in,height=1.6in,angle=-90]{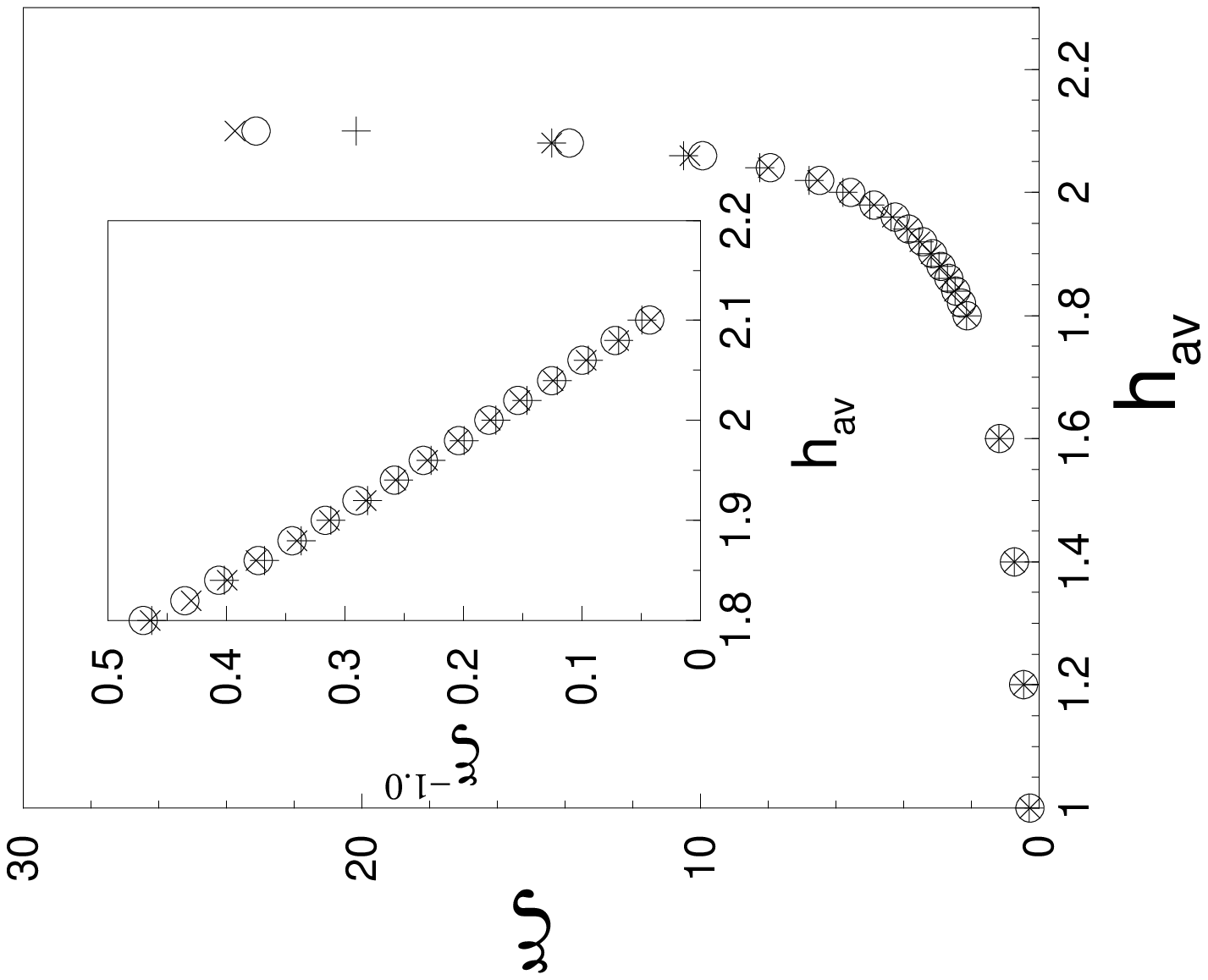}\hskip.1in\includegraphics[width=2.2in,height=1.6in,angle=-90]{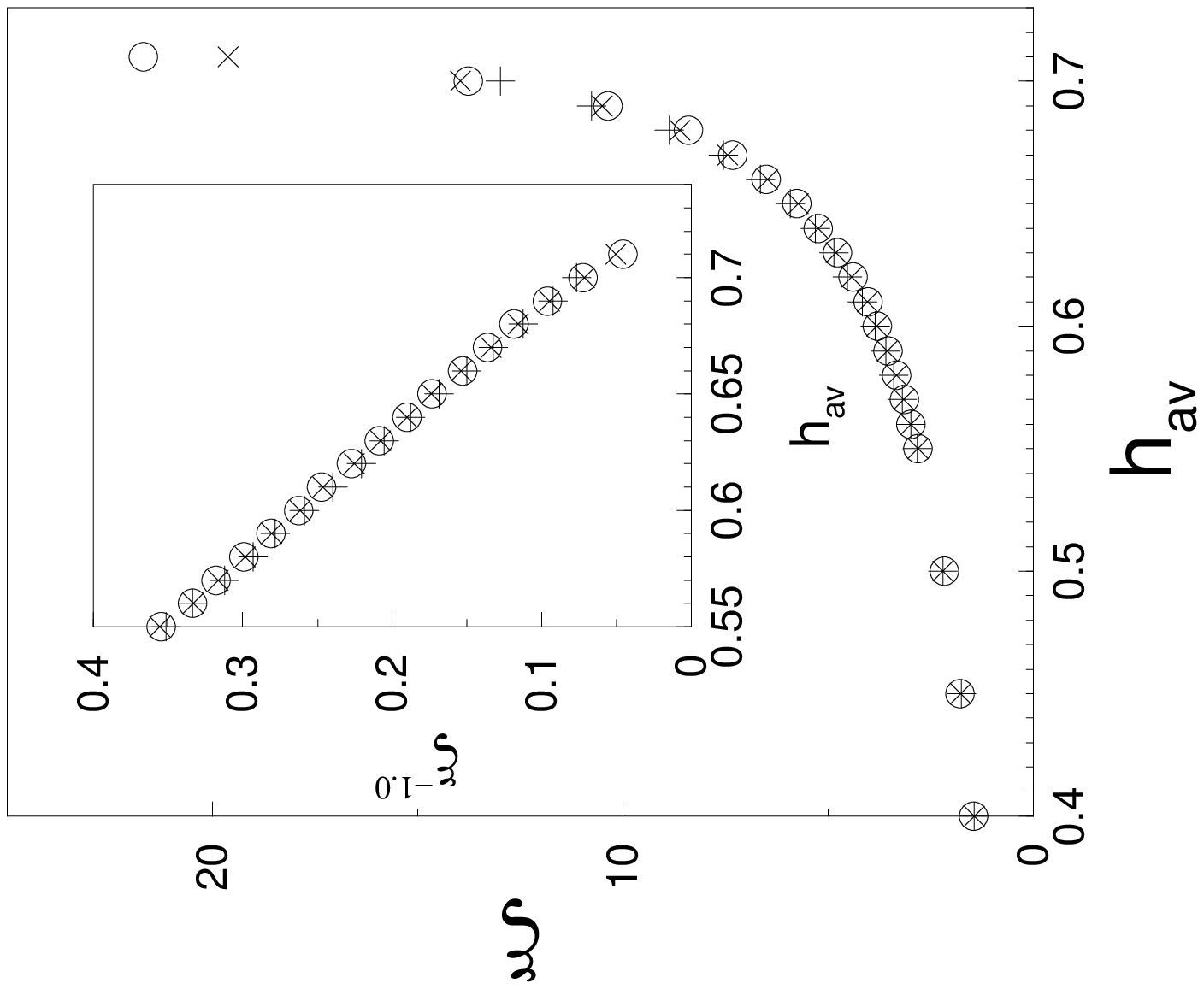}\par\end{flushleft}

{\footnotesize Figure} \textbf{\footnotesize 13}{\footnotesize :}
\textbf{\footnotesize }{\footnotesize Sub-critical response: Correlation
length ($\xi$) versus average hight ($h_{av}$) in BTW model (left)
and in Manna model (right). Inset shows the plot of inverse $\xi$
versus $h_{av}$ and predicts the critical hight through extrapolation.}{\footnotesize \par}

\vskip.1in

Therefore, although BTW and Manna models belong to different universality
classes with respect to their properties at the critical state, both
the models show similar sub-critical response or precursors. A proper
extrapolation method can estimate the respective critical heights
of the models quite accurately. 

\vskip.1in

\begin{flushleft}\textbf{(D)} \textbf{\emph{Fractal models}}\par\end{flushleft}

\vskip.1in

The surfaces of earth's crust and tectonic plate at the fault zone
are not compact rather fractal in nature. In fact, these surfaces
are the results of the large scale fracture separating the crust from
the moving tectonic plate. It has been observed that these surfaces
are self similar fractals \cite{BS..} having the self-affine scaling
property $h(\lambda x,\lambda y)\sim\lambda^{\zeta}h(x,y)$, where
$h(x)$ denotes the height of the crack surface at the point $x$
and $\zeta$ is the roughness exponent. It has been claimed recently
that since the fractured surfaces have got well-characterized self-affine
properties, the distribution of the elastic energies released during
the slips (earthquake events) between two rough surfaces (crust and
plate) may follow the overlap distribution of two fractal surfaces
\cite{V96,CS99}

\begin{center}\includegraphics[width=3in,height=1.5in]{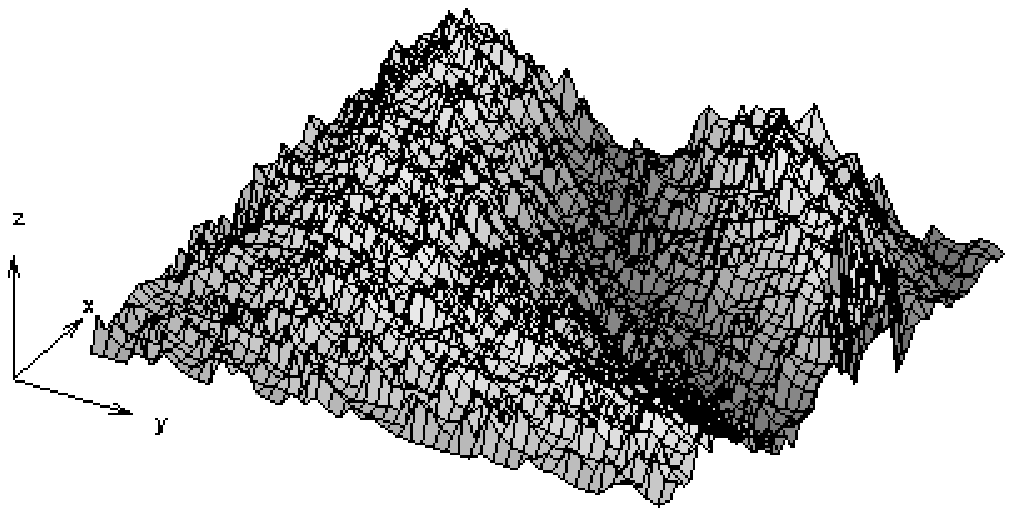}\par\end{center}

{\footnotesize Figure} \textbf{\footnotesize 14}{\footnotesize : A
typical self-affine fracture surface} \textbf{\footnotesize }{\footnotesize in
($2+1$) dimension with roughness exponent $\zeta=0.8$.}{\footnotesize \par}

\subsection*{\textmd{\emph{Self-affine asperity model }}}

\noindent V. De Rubies et. al. \cite{V96} has proposed a new model
for earthquakes where the scale invariance of the Gutenberg-Richter
law is claimed to come from the fractal geometry of the fault surfaces.
In this model the sliding fault surfaces have been represented by
fractional Brownian surfaces, whose height scales as $|h(x+r)-h(x)|\sim r^{\zeta}$.
The roughness of the surfaces are determined from the value of the
exponent $\zeta$ which lies between $0$ and $1$. Two surfaces are
simulated by two statistically self-affine profiles, say, $h_{1}(x)$
and $h_{2}(x)$, one drifting over other with a constant speed $v$
such that $h_{1}(x,t)=h_{2}(x-vt)$. An interaction between two profiles
represents a single seismic event and the energy released is assumed
to be proportional to the breaking area of the asperities. At the
contact point of the surfaces the `surface roughness' prevents slipping
and the stress is accumulated there. When the stress exceeds a certain
threshold value, breaking (earthquake) occurs. This model produces
Gutenberg-Richter type power law: $P(E)\sim E^{-\beta-1}$, where
$P(E)dE$ is the probability that an earthquake releases energy between
$E$ and $E+dE$ and the exponent $\beta$ is directly related to
the roughness exponent $\zeta$ as : $\beta=1-\zeta/(d-1)=(D_{f}-1)/(d-1)$,
where $D_{f}$ and $d$ are respectively the fractal dimension and
the embedding dimension of the surfaces.

\subsection*{\textmd{\emph{Chakrabarti-Stinchcombe model }}}

\noindent This is an analytical model, proposed by Chakrabarti and
Stinchcombe \cite{CS99}, which incorporates the self-similar nature
of both the crust and the tectonic plate. They used self-similar fractals
to represent fault surfaces (Fig.15).

\begin{center}\includegraphics[%
  width=3in,
  height=1.5in]{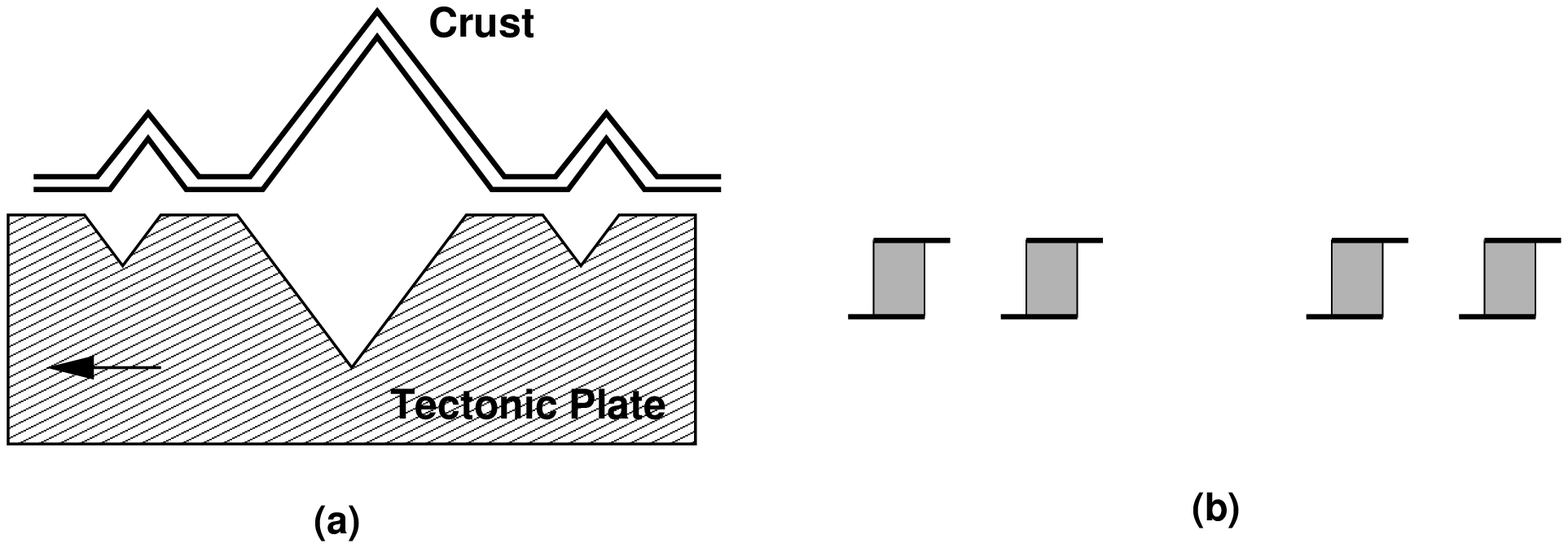}\par\end{center}

{\footnotesize Figure} \textbf{\footnotesize 15}{\footnotesize :}
\textbf{\footnotesize }{\footnotesize Schematic representation of
the rough surfaces of earth's crust and moving tectonic plate \cite{CS99}. }{\footnotesize \par}

\vskip.1in

\noindent The total contact area between the surfaces is assumed to
be proportional to the elastic strain energy that can be grown during
the sticking period, as the solid-solid friction force arises from
the elastic strain at the contacts between the asperities. This energy
is considered to be released as one surface slips over the other and
sticks again to the next contact between the rough surfaces. Chakrabarti
and Stichcombe have shown analytically through renormalization group
calculations that for regular fractals (Cantor sets and carpets) the
contact area follows power law distribution: \[
\rho(s)\sim s^{-\gamma};\gamma=1,\]
which is comparable to that of Gutenberg-Richter law. 

\vskip.1in

\subsection*{\textmd{\emph{Numerical verification}}}

The claim of Chakrabarti-Stichcombe model has been verified by extensive
numerical simulations \cite{PCRD-03} taking different type of synthetic
fractals: regular or non-random Cantor sets, random Cantor sets (in
one dimension), regular and random gaskets on square lattice and percolating
clusters embedded in two dimensions. 

\begin{center}\includegraphics[width=1.5in,height=1.5in]{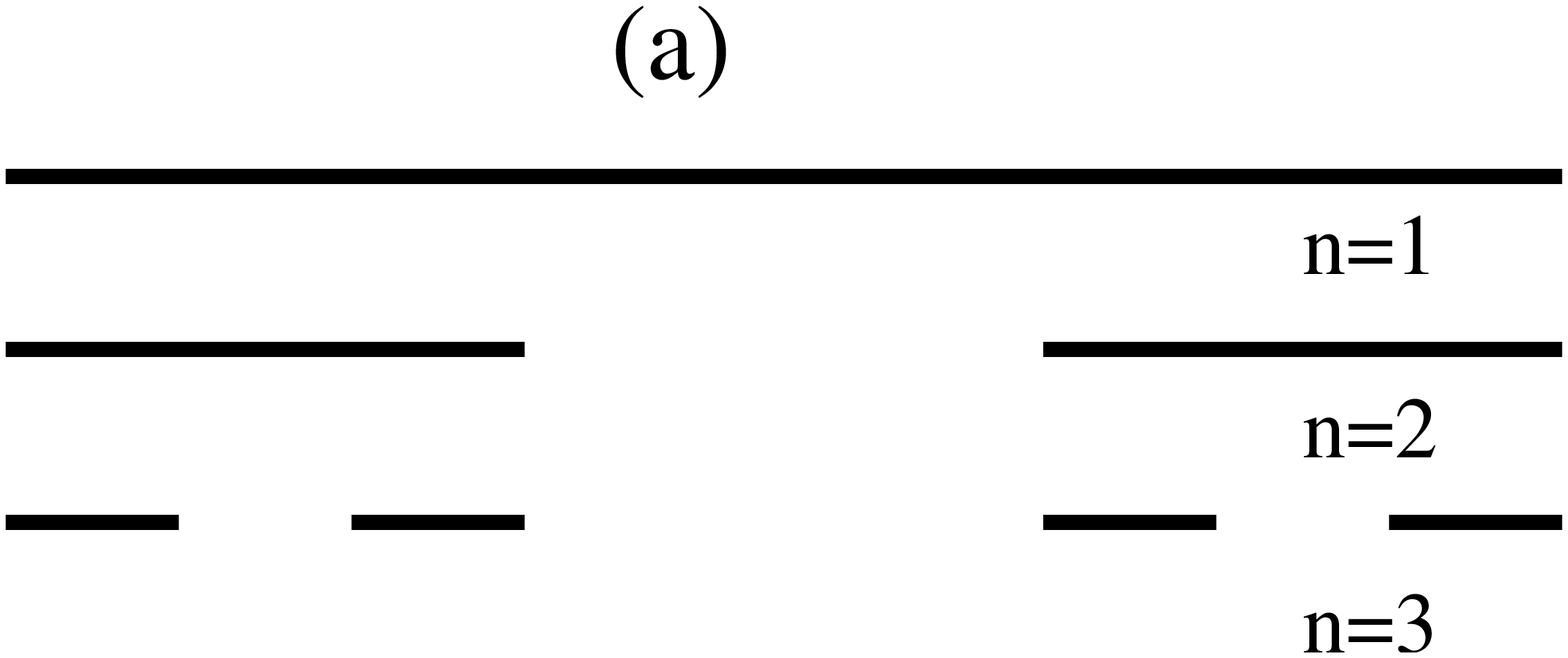}\hskip.2in\includegraphics[width=1.5in,height=1.5in]{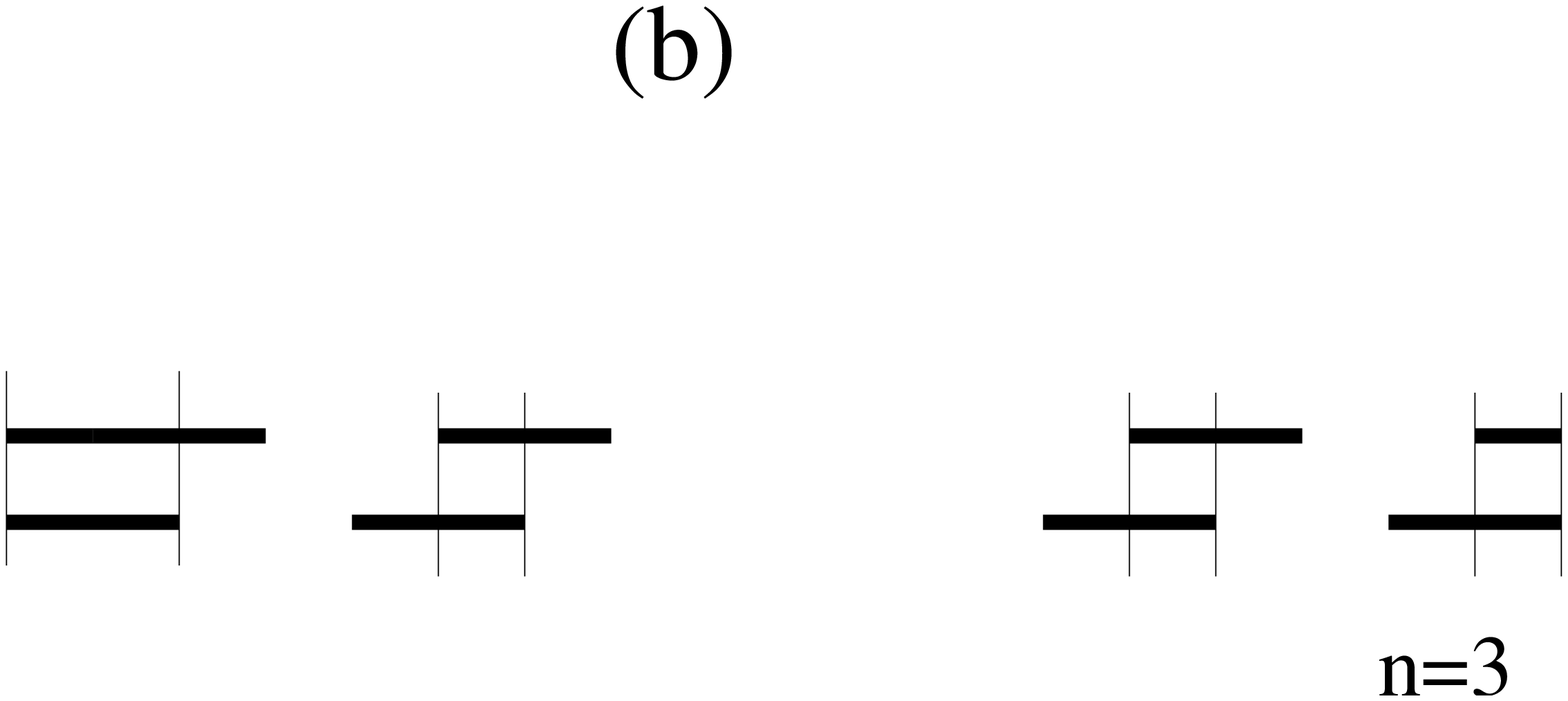}\par\end{center}

\vskip.1in

{\footnotesize Figure} \textbf{\footnotesize 16}{\footnotesize :}
\textbf{\footnotesize }{\footnotesize (a) A regular Cantor set of
dimension $\ln2/\ln3$; only three finite generations are shown. (b)
The overlap of two identical (regular) Cantor sets, at $n=3$, when
one slips over other; the overlap sets are indicated within the vertical
lines, where periodic boundary condition has been used. }{\footnotesize \par}

\vskip.4in

The contact area distributions $P(m,L)$ seem to follow a universal
scaling: \[
P(m,L)\sim L^{\alpha}P^{\prime}(m^{\prime});m^{\prime}=mL^{\alpha},\]

\noindent where $L$ denotes the size of the fractal and $\alpha=2(d-d_{f})$;
$d_{f}$ being the mass dimension of the fractal and $d$ is the embedding
dimension. Also the overlap distribution $P(m)$, and hence the scaled
distribution $P^{\prime}(m^{\prime})$, decay with $m$ or $m^{\prime}$
following a power law (Fig.17) for both regular and random Cantor
sets and gaskets: \[
P(m)=m^{-\beta};\beta=d.\]

A very recent report \cite{pratip05} analytically explains the origin
of such asymptotic power laws in case of Cantor set overlap.

\begin{center}\includegraphics[width=2in,height=2.5in,angle=-90]{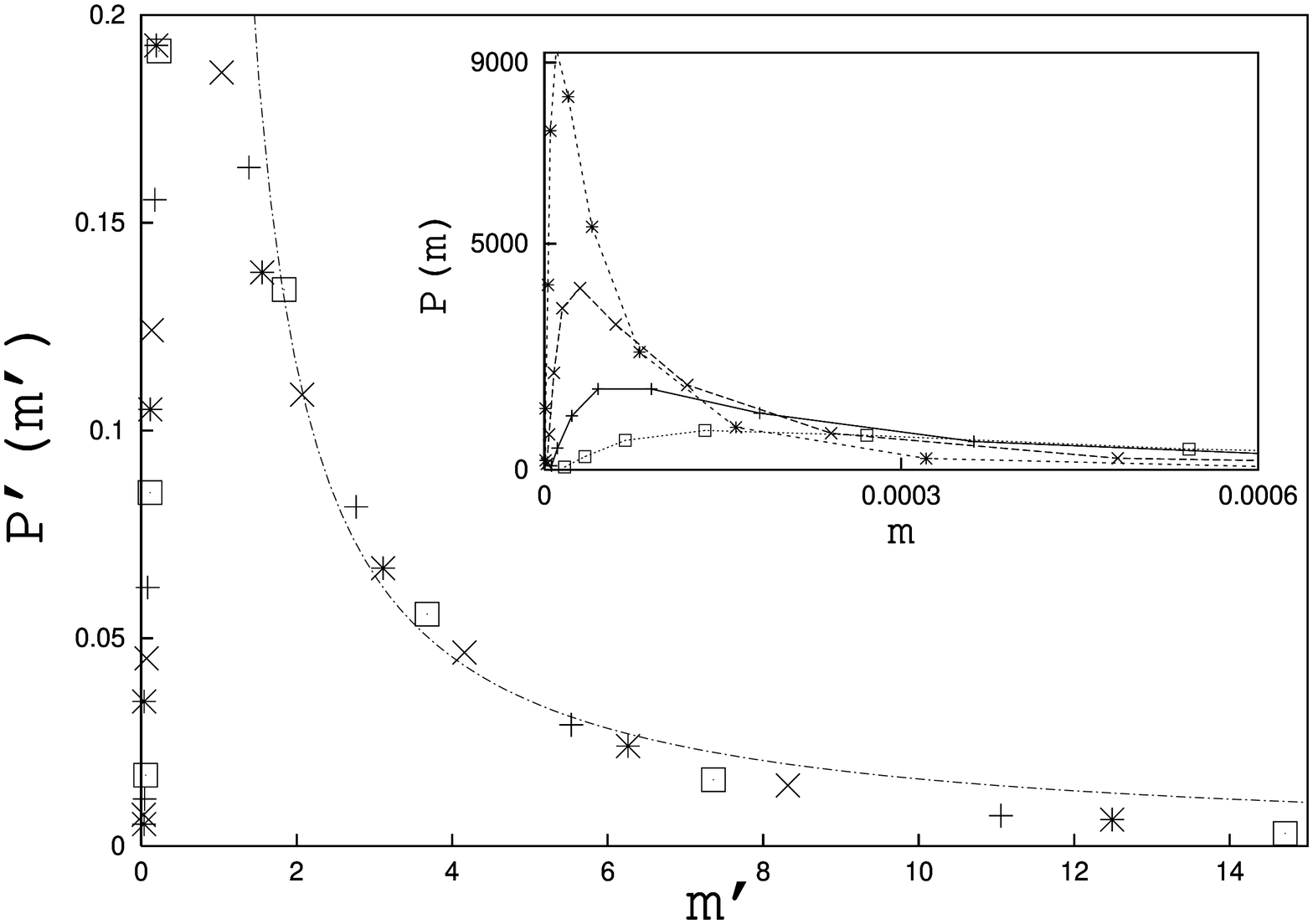}\par\end{center}

\vskip.1in

{\footnotesize Figure} \textbf{\footnotesize 17}{\footnotesize : The
plot of $P^{\prime}(m^{\prime})$ against $m^{\prime}$ for Cantor
sets with $d_{f}=\ln2/\ln3$ at finite generations: $n=10$ (square),
$n=11$ (plus), $n=12$ (cross) and $n=13$ (star) and the dotted
lines indicate the best fit curves of the form $a(x-b)^{-d}$; where
$d=1$. Inset shows $P(m)$ vs. $m$ plots. }{\footnotesize \par}

\subsection*{\textmd{\emph{Prediction possibility of large events}}}

If one cantor set moves uniformly over other, the overlap between
the two fractals change quasi-randomly with time and produces a time
series of overlaps $m(t)$. Such a time series is shown in Fig.18,
for Cantor sets of dimensions $\ln2/\ln3$. While most of the overlaps
are small in magnitude, some are really big, where as the cumulative
overlap size $Q(t)=\int_{o}^{t}mdt$ `on average' grows linearly with
time. Is it possible to predict a large future overlap analysing the
time series data? A recent study \cite{PCC} suggests a method: 

\begin{center}\includegraphics[width=2.5in,height=2.2in]{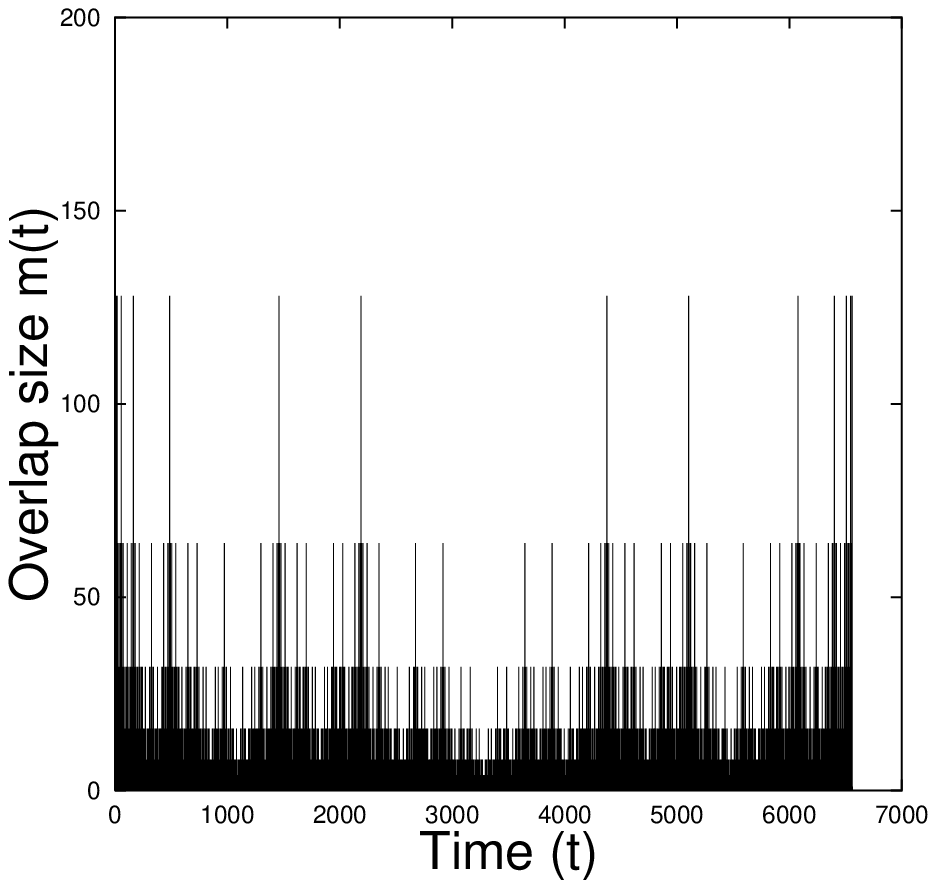}\par\end{center}

{\small Figure} \textbf{\small 18}{\small :} \textbf{\small }{\small The
time ($t$) series data of overlap size ($m$) for regular Cantor
sets: of dimension $\ln2/\ln3$, at $8$th generation. }{\small \par}

\vskip.1in

One can identify the `large events' occurring at time $t_{i}$ in
the $m(t)$ series, where $m(t_{i})\geq M$, a pre-assigned number,
then calculate the cumulative overlap size $Q(t)=\int_{t_{i}}^{t_{i+1}}mdt$,
where the successive large events occur at times $t_{i}$ and $t_{i+1}$.
Obviously $Q(t)$ is reset to $0$ value after every large event.
The behavior of $Q_{i}$ with time is shown in Fig.19 for regular
cantor sets . It appears that there are discrete values up to which
$Q_{i}$ grows with time.

\vskip.1in

\begin{center}\includegraphics[width=1.5in,height=1.7in]{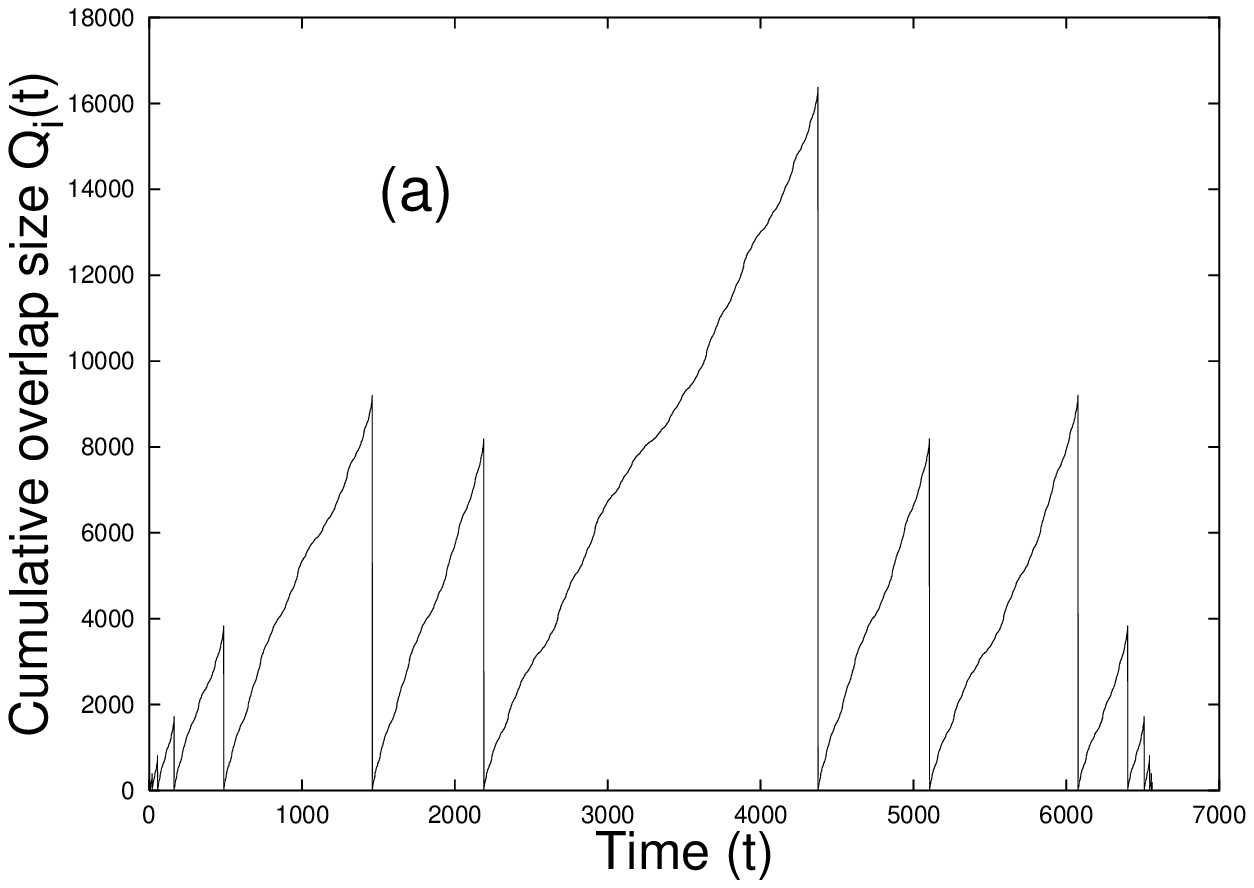}\includegraphics[width=1.5in,height=1.7in]{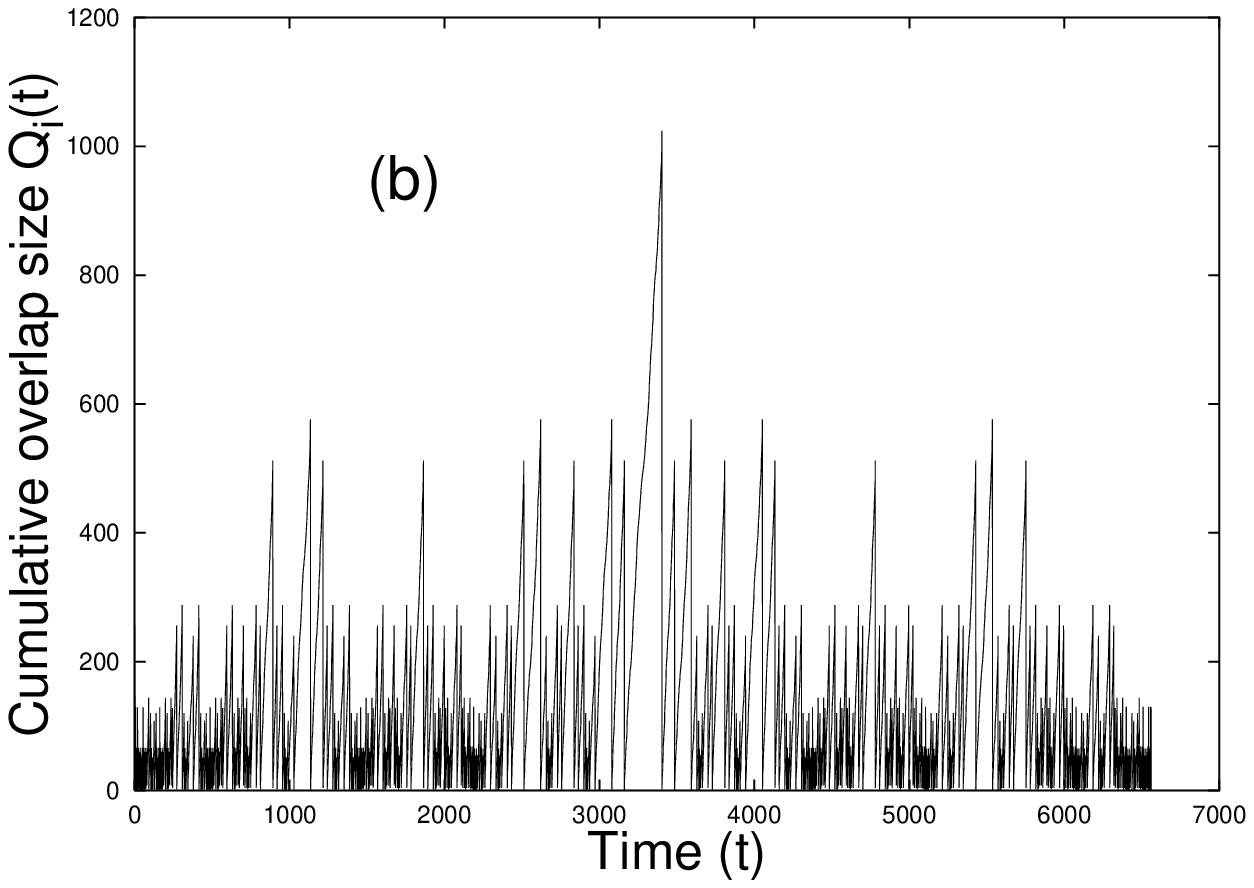}\par\end{center}

\vskip.1in

{\small Figure} \textbf{\small 19}{\small : The cumulative overlap
size variation with time (for regular Cantor sets of dimension $\ln2/\ln3$,
at $8$th generation), where the cumulative overlap has been reset
to $0$ value after every big event (of overlap size $\geq M$ where
$M=128$ and $32$ respectively).} 

\vskip.6in

Therefore, if one fixes a magnitude $M$ of the overlap sizes $m$,
so that overlaps with $m\geq M$ are called `events' (or earthquake),
then the cumulative overlap $Q_{i}$ grows linearly with time up to
some discrete levels $Q_{i}\cong lQ_{0}$, where $Q_{0}$ is the minimal
overlap size, dependent on $M$ and $l$ is an integer. This information
certainly does not help to predict a future large event accurately,
but it gives some hints by identifying discrete levels of $Q_{i}$
where a large overlap is likely to happen.

\vskip.1in

\begin{flushleft}\textbf{\emph{Discussions and concluding remarks }}\par\end{flushleft}

\vskip.1in

The Gutenberg-Richter law is a well established law in earthquake
research. It should be mentioned that Gutenberg and Richter obtained
this law from the statistics of earthquake events observed throughout
the world. Also, earthquakes within a tectonically active region (Japan,
California etc) follow similar power law. It is still a controversial
issue whether the exponent of the Gutenberg-Richter power law is an
universal constant or it varies in a narrow range ($0.8$ to $1.2$)
depending upon the nature of the fault zone. As the motion of the
tectonic plates is surely an observed fact, the stick-slip process
should be a major ingredient of earthquake models. Although Burridge-Knopoff
type spring-block model successfully captures the stick-slip dynamics
and reproduces Gutenberg-Richter type power law in the size distribution
of events, it is far from the accurate representation of earthquake
dynamics -as it does not contain a mechanism for aftershocks which
are observed facts. But this model has some important features: The
dynamics is inherently chaotic -therefore technically unpredictable
which agrees well with the occurrence of earthquakes. However, the
method proposed by Dey et. al \cite{rumi} to predict a major slip
event by monitoring the cumulative energy function, may open up a
wide scope of future research in this field. On the other hand, fiber
bundle model and fuse model have been developed basically to study
breakdown phenomena in composite materials. As earthquake is a major
breakdown phenomenon, these models can be used as earthquake models
due to their inherent mean-field nature. Avalanche distributions show
power laws in both the models and a crossover in exponent value appears
\cite{PHH-05} near breakdown point, which can be treated as a criterion
for imminent breakdown. SOC models assume a self-driven slow dynamics
and reproduces Gutenberg-Richter law at the critical point. Although
the critical state can be predicted accurately from the sub-critical
response of the systems, the behavior remains unpredictable at the
critical state. Fractal overlap models are different (from the other
three types) modelling approaches in the sense that they focus on
the fractal nature of the fault interfaces and not on the dynamics.
The contact area distribution follows asymptotic power law and this
suggests a possibility that fractal geometry of the faults might be
the true origin of Gutenberg-Richter law. 

There are several difficulties in studying earthquake phenomenon,
as it is a ``N body'' complex problem. While exact solution of a
$3$-body problem needs rigorous mathematical calculations and it
does not work for a mutually interacting system having more than $3$
bodies, naturally, theoretical physics can not help to formulate the
entire earthquake dynamics. Again, as the dynamics of earthquake happens
at a depth more than $20$ km from earth surface -experimental observation
of such dynamics is almost impossible. Moreover, earthquake dynamics
involves crust and plates which are highly heterogeneous at many scales
from the atomic scale to the scale of tectonic plates, with the presence
of dislocation, impurities, grains, water etc and the scenario becomes
very much complicated. In this situation, model studies are very important
in earthquake research in the sense that they can produce synthetic
earthquake events, allow us to monitor the dynamics and analyse the
event statistics to compare with the real earthquake data. Moreover,
such studies suggest potential methods to predict a major event. It
will be a real breakthrough if any of such methods can help a little
to predict a future earthquake. 

\vskip.1in

\begin{flushleft}\textbf{\emph{Acknowledgment: }}\par\end{flushleft}

\vskip.1in

We are grateful to Bikas. K. Chakrabarti for important collaborations
and useful suggestions. Thanks to Research Council of Norway (NFR)
for financial support through grant No. 166720/V30.

\end{document}